\documentclass[aps,prl,twocolumn,showpacs,amsmath,amssymb]{revtex4}

\usepackage{graphicx}
\usepackage{color}

\definecolor{Red}{rgb}{1.0,0.0,0.0}

\def\be#1{\begin{equation}\label{#1}}
\def\ee{\end{equation}}

\begin{document}

\title{Non-minimal couplings in Randall-Sundrum Scenarios}

\author{G. Alencar${}^a$}
\email{geova@fisica.ufc.br}
\author{R. R. Landim${}^a$}
\email{rrlandim@gmail.com}
\author{C. R. Muniz${}^{b,}$}
\email{celiomuniz@yahoo.com}
\author{R.N. Costa Filho${}^a$}
\email{rai@fisica.ufc.br}
\affiliation{${}^a$Departamento de F\'{\i}sica, Universidade Federal
do Cear\'a, 60451-970 Fortaleza, Cear\'a, Brazil}
\affiliation{${}^b$Universidade Estadual do Cear\'a, Faculdade de Educa\c c\~ao, Ci\^encias e Letras do Sert\~ao Central -
R. Epit\'acio Pessoa, 2554, 63.900-000  Quixad\'{a}, Cear\'{a},  Brazil.}

\date{\today}

\begin{abstract}

  In this paper we propose a new way of obtaining four dimensional gauge invariant $U(n)$ gauge field from a bulk action. The results are valid for both Randall-Sundrum scenarios and are obtained without the introduction of other fields or new degrees of freedom. The model is based only in non-minimal couplings with the gravity field. We show that two non-minimal couplings are necessary, one with the field strength and the other with a mass term. Despite the loosing of five dimensional gauge invariance by the mass term a massless gauge field is obtained over the brane. To obtain this, we need of a fine tuning of the two parameters introduced through the couplings. The fine tuning is obtained by imposing the boundary conditions and to guarantee non-abelian gauge invariance in four dimensions. With this we are left with no free parameters and the model is completely determined. The model also provides analytical solutions to the linearized equations for the zero mode and for a general warp factor.

\end{abstract}

\maketitle

The Randall Sundrum (RS) model appeared in the Physics of higher dimensions
as an alternative to compactification that included the possibility
of solving the Hierarchy problem \cite{Randall:1999ee,Randall:1999vf}. To solve the physical problem of
dimensional reduction, the RS models should obtain fields with zero
mode confined to the brane in order to recoverer the Physical models
when the fields are properly integrated over the extra dimension.
In the non-compact case, the gravity field localization is attained, however gauge fields as simple as $U(1)$ minimally coupled to gravity are not localized
and this was a problem to the theory\cite{Randall:1999vf,Bajc:1999mh}. For the compact case the problem appears when we consider non-abelian gauge fields. After the Fourier decomposition of the zero mode and the normalization $\int dz \psi^{2}=1$ we have that $\int dz \psi^{3} _{m=0}\neq 1$, and the gauge invariance in four dimensions is lost~\cite{Batell:2006dp}. A number of extensions to RS were proposed in order to provide localized gauge fields for the non-compact case.
A smooth warp factor was investigated?\cite{Batell:2006dp,Dvali:1996xe,Gremm:1999pj,Kehagias:2000au,Bazeia:2004dh,Cvetic:2008gu,Chumbes:2011zt,German:2012rv},
but also it did not yield localized gauge fields.
Some models obtained the localization by the addition of new degrees of freedom
such as a scalar or dilaton fields, but a more natural approach would be to obtain an extension
of minimal couplings that would localize the gauge field without introducing
other fields. A step in this direction was the introduction of a boundary interaction with the field strength \cite{Dvali:2000rx}, but it is known that this produces
only a quasi-localized zero mode.

Other idea is the breaking of the $U(1)$ gauge invariance in five dimensions
by the addition of a mass term. It was found that the only consistent way of getting
a localized zero mode is again the introduction of a boundary term, this time with the
mass term~\cite{Ghoroku:2001zu}. Soon it was shown that the boundary mass term was
not enough to provide a consistent model in the case of Yang-Mills
(YM) fields because non-abelian gauge invariance is lost in the membrane~\cite{Batell:2006dp}.
The work~\cite{Batell:2006dp} shows that a combination of the above boundary
terms are needed to solve the problem of YM localization.  In the track of understanding
the origin of the mass boundary term introduced in~\cite{Ghoroku:2001zu}, some of us identified
that a non-minimal coupling of the Ricci scalar with the mass term
could generate it~\cite{Alencar:2014moa}. More then this, it was discovered that this
kind of coupling solves the localization of any $U(1)$ $p-$form gauge
field in co-dimension one brane worlds, valid for any warp factor~\cite{Alencar:2014fga}. However the mechanism
does not work for Yang-Mills fields by the same reasons of Ref.~\cite{Batell:2006dp}. In fact, as shown by these authors, this is a problem even for the non-compact case since non-abelian gauge symmetry is lost over the brane if we depart from a massless gauge invariant action in the bulk.
Other non-minimal couplings of gravity with the field strength seems
to be needed also in this case. These kinds of couplings of gauge fields
with gravity has been proposed and studied in 4D~\cite{Horndeski:1976gi} and recently it was shown that they generate the
boundary term for the field strength in Ref.~\cite{Germani:2011cv}.
However the model keeps the property of being
only quasi-localized. Here we show that specific non-minimal couplings with
gravity are enough to recover YM fields over the brane, moreover,
the confined fields displays the expected properties, that is, are massless
and therefore gauge invariant. Our approach also has the advantage of being
valid for any extension of the RS
model which recovers it asymptotically.
This is obtained when we merge the coupling as
proposed in~\cite{Horndeski:1976gi} with the non-minimal coupling proposed
by us in Ref.~\cite{Alencar:2014moa}. Importantly, localization and gauge invariance
depend on the tuning of two otherwise free parameters of the model.

The background metric of the RS model in its conformal form is given
by $ds^{2}=e^{2A(z)}\eta_{MN}dx^{M}dx^{N}$, where $z=x^{5}$. After
considering a delta-like branes and an AdS vacuum a stable background
solution to the Einstein equation is found, which is given by $A=-\ln(k|z|+1)$.
The question of localizability of fields is resumed to find a square
integrable solution($\psi$) of a Schroedinger-like mass equation
with potential $U(z)$ which emerges from the process of dimensional
reduction. At the end this provides a finite four dimensional action
after the integration in the extra dimension $\int\psi^{2}dz$ is
performed. For the non-abelian case the problem is reduced to finding a zero mode satisfying $\int dz \psi^{2}=\int dz \psi^{3}$.
The above warp factor generates effective potentials with
Dirac delta functions~\cite{Randall:1999ee,Randall:1999vf}. These singularities can be smoothed out through
the introduction of kink-like membranes such as in Refs. \cite{Dvali:1996xe,Gremm:1999pj,Kehagias:2000au,Bazeia:2004dh,Cvetic:2008gu}  that recovers the RS case only asymptotically.
Therefore the best approach is to consider a general warp factor $A$ and to look for a solution to the problem that does not depend on it.
We call such solution warp independent
for obvious reasons. For a minimally coupled $U(1)$ gauge field the
effective potential obtained is $U=A''/2+A'^{2}/4$. Analyzing more
carefully, for a thin brane, the solution has a convergent and a divergent
component as $z\to\infty$. However this is not the only boundary
condition to be imposed since $A$ depends on $|z|$ and we get a
delta function in the potential. When we impose the proper boundary condition
in $z=0$ the solution is completely fixed and we discover that it
is not localized. The same conclusion is obtained for arbitrary
$A$ and it is warp independent. The form of the above potential provides
a general solution given by $\psi=e^{-A/2}$ and this is not square
integrable for asymptotic RS models. Also, this solution does not satisfy the condition $\int dz \psi^{2}=\int dz \psi^{3}$ needed for the $U(n)$ compact case.

As said before, a model was constructed which introduces
a boundary term through an interaction with a delta function~\cite{Dvali:2000rx}, that is,
besides the standard gauge action a contribution given by $\frac{\delta(z)}{m}F_{\mu\nu}^{2}$
was introduced, where $\mu$ are four dimensional indices and $m$
is a mass parameter. Although the origin of this term is not
explained, the authors argue that it is needed to guarantee that there
is no current in the extra dimension. Then a quasi-localized zero
mode is obtained. Some time latter the possible gravitational origin
of such an interaction was discovered to come from a non-minimal coupling
with the field strength given by
$\Delta^{ABCD}F_{AB}F_{CD}$~\cite{Germani:2011cv}. This coupling has been proposed early as an extension of the
four dimension gauge field which preserve current conservation and
recovers Maxwell equations in the flat limit~\cite{Horndeski:1976gi}. This is obtained
if $\nabla_{A}\Delta^{...A...}=0$ for any of the indices and the tensor
posses the same symmetries of the curvature tensor. A $\Delta$ satisfying this is given by
\begin{equation} \label{delta}
\Delta^{AB}{}_{CD}\equiv\frac{1}{8}R^{AB}{}_{CD}-\frac{1}{2}R^{[A}{}_{[C}\delta^{B]}{}_{D]}+\frac{1}{8}R\delta^{[A}{}_{[C}\delta^{B]}{}_{D]}.
\end{equation}

By computing explicitly the above coupling for the thin brane
case Germani has obtained the correct boundary term~\cite{Germani:2011cv}. Despite this
clear advance the model does not provide a fully localized solution.

In the direction of five dimension breaking of gauge invariance a
topological mass term was added through a three form field with a
localized zero mode~\cite{Oda:2001ux}. However this generates a massive gauge
field in four dimension, beyond introducing a form field as a new
degree of freedom. To solve
this problem Ghoroku et al proposed to introduce a mass $M^{2}A_{N}^{2}$
term directly in the action~\cite{Ghoroku:2001zu}. Although this term generates a localizable
solution as $z\to\infty$ the boundary condition at $z=0$ fixes $M=0$.
This is solved in a similar way as in Ref.\cite{Dvali:2000rx} by the introduction of
a boundary mass term $\delta(z)A_{N}^{2}$ such that the boundary
condition is satisfied with $M\neq0$. Therefore the model is consistent
if we use a interaction term given by $M(z)=(a+b\delta(z))$
where $a$ and $b$ should be chosen such that the boundary condition
are satisfied and to obtain a localized solution. A range in the parameters
is obtained which satisfy the boundary conditions leaving a free parameter to be determined.
However it was shown in \cite{Batell:2006dp} that
this is not enough when we consider YM fields. The reason is that
now we have quartic terms in the action and even if the zero mode
is integrable we generally have $\int\psi^{3}dz\neq\int\psi^{4}dz$
causing the lost of gauge invariance after the integration is performed.
However Batell et al showed that this can be solved by the use of both
boundary terms of Refs. \cite{Dvali:2000rx,Ghoroku:2001zu} and the final action which was found to be
\begin{eqnarray}
S&=&-\frac{1}{4e_{5}^{2}}\int d^{5}x\sqrt{-g}\left[TrF_{MN}^{a}F^{MNa}+\beta\delta(z)TrF_{\mu\nu}^{a}F^{\mu\nu a}\right]\nonumber \\
&-&\frac{1}{2}\int d^{5}x\sqrt{-g}M^{2}(z)TrA_{M}^{a}A^{Ma}
\end{eqnarray}
where $e_{5}$ is the five dimensional gauge coupling and $F_{MN}^{a}=\partial_{M}A_{N}^{a}-\partial_{N}A_{M}^{a}+f^{abc}A_{M}^{b}A_{N}^{c}$.
Since $M(z)=(a+b\delta(z))$ the model posses three parameters: $a$ and $b$ to guarantee the localization and a third parameter $\beta$ is introduced to guarantee the gauge invariance
in four dimensions. However, we show here that this action must be corrected to preserve the expected symmetries of the system.

Searching for the origin of the above action, some of us discovered
in a series of previous papers that a non-minimal coupling with the
Ricci scalar given by $\gamma RA_{N}^{2}$ consistently provides a
solution to the localization of $U(1)$ gauge field in co-dimension
one brane worlds\cite{Alencar:2014moa,Alencar:2014fga,Jardim:2014vba}. This term modifies the potential that now is given
by $U=pA''+p^{2}A'^{2}$ where $p$ is the order of the $p-$form
considered. The zero mode for this potential has solution $\psi=e^{pA}$
which is square integrable for any warp factor that asymptotically
recovers RS, being therefore a warp independent solution. Another
advantage is that the exigence of general covariance determines
one of the parameters leaving only $\gamma$ which is fixed by boundary
conditions, and at the end we get a model with no free parameters.
However the problem for YM fields is not solved and other non-minimal
coupling seems to be needed.

With the above ingredients we propose a model defined by the following action
\begin{eqnarray}\label{action}
S&=&-\frac{1}{4e_{5}^{2}}\int d^{5}x\sqrt{-g}\left[TrF_{MN}^{a}F^{MNa}+\gamma_{1}\Delta^{AB}{}_{CD}TrF_{AB}^{a}F^{CDa}\right] \nonumber \\
&-&\frac{\gamma_{2}}{2}\int d^{5}x\sqrt{-g}RTrA_{M}^{a}A^{Ma},
\end{eqnarray}
where $\gamma_{1}$ and $\gamma_{2}$ are parameters that will be
fixed by the boundary condition and the demand of gauge invariance in $4D$.
From now on we will not write explicitly the group indices. The linearized equations
of motion are
\begin{equation}\label{EOM}
\partial_{M}(\sqrt{g}F^{MN})+\gamma_{1}\partial_{M}(\Delta^{MNOP}\sqrt{g}F_{OP})=-\gamma_{2}R\sqrt{g}g^{MN}A_{M}
\end{equation}
and from the above we get the five dimensional divergenceless condition
$\partial_{M}(\sqrt{g}Rg^{MN}A_{N})=0$, or in components
\begin{equation}\label{divergence}
\partial_{5}(Re^{3A}\Phi)+Re^{3A}\partial_{\mu}A^{\mu}=0,
\end{equation}
where we have defined $\Phi\equiv A^{5}$. By computing $\Delta^{A...}$
explicitly we obtain the equations of motion in components. For $N=\nu$
we get
\begin{equation}\label{nu}
e^{A}(1+\gamma_{1}f)\partial_{\mu}F^{\mu\nu}+\partial(e^{A}(1+\gamma_{1}h)F^{5\nu})=-\gamma_{2}Re^{3A}A^{\nu}
\end{equation}
and for $N=5$
\begin{equation}\label{5}
e^{A}(1+\gamma_{1}h(z))\partial_{\mu}F^{\mu5}=-\gamma_{2}Re^{3A}\Phi
\end{equation}
where $e^{2A}h=-3A'^{2}/4$ and $e^{2A}f=-A''/2-A'^{2}/4$. Therefore
our system is defined by the three equations
above, which include
a vector and a scalar field in four dimensions. From Eq.(\ref{divergence}) we
see that our four dimensional vector field does not satisfy the four
dimensional divergenceless condition. The strategy, as explained before
in Ref. \cite{Alencar:2014moa} is to split the field in longitudinal and transverse
parts $A_{T}^{\mu}=(\delta_{\nu}^{\mu}-\frac{\partial^{\mu}\partial_{\nu}}{\Box})A^{\nu},A_{L}^{\mu}=\frac{\partial^{\mu}\partial_{\nu}}{\Box}A^{\nu}$
and to interpret(\ref{divergence}) as an equation relating the longitudinal part
of $A^{\mu}$ and the scalar field. The main question posed at this point is if we can decouple the transverse component in order to obtain
a well defined YM field in four dimensions. For the $U(1)$ case and for
$\gamma_{1}=0$, with only the coupling to the mass term,
we have shown that this is the case. Now we have the additional complication
of the non-minimal coupling with the field strength, however we will
see that it is also valid for the present case. For this end first
we define $F_{L}^{5\mu}=\partial^{5}A_{L}^{\mu}-\partial^{\mu}\Phi$,
from where we get the identity $F_{L}^{5\mu}=-\frac{\partial^{\mu}}{\Box}\partial_{\nu}F^{\nu5}$.
Beyond this we also have $\partial_{\mu}F^{\mu\nu}=\Box A_{T}^{\nu}$
and Eq. (\ref{nu}) becomes
\begin{eqnarray}
&&e^{A}(1+\gamma_{1}f)\Box A_{T}^{\nu}+\partial(e^{A}(1+h\gamma_{1})A_{T}^{\nu})+\gamma_{2}Re^{3A}A_{T}^{\nu}+\nonumber \\
&&\partial(e^{A}(1+h\gamma_{1})F_{L}^{5\nu})+\gamma_{2}Re^{3A}A_{L}^{\nu}=0.
\end{eqnarray}

Now by using Eq. (\ref{5}), from the definition of $A^{\mu}_L$ and our identity for $F_{L}^{5\mu}$
we can show that
\[
\partial(e^{A}(1+h\gamma_{1})F_{L}^{5\nu})=-\gamma_{2}Re^{3A}A_{L}^{\nu}
\]
where we have used the fact $\partial_5 h=0$ from the tracelessness of $\Delta$. Therefore we successfully get our decoupled equation for the transverse
YM field given by
\[
e^{A}(1+\gamma_{1}f)\Box A_{T}^{\nu}+\partial(e^{A}(1+h\gamma_{1})\partial A_{T}^{\nu})+\gamma_{2}Re^{3A}A_{T}^{\nu}=0.
\]

Now performing the standard separations of variables $A_{T}^{\mu}=\tilde{A}_{T}^{\mu}(x^{\mu})\theta^{-\frac{1}{2}}(z)\psi(z)$,
with $\theta=e^{A}(1+h\gamma_{1})$ and by considering the zero mode $\Box\tilde{A}_{T}^{\mu}=0$
we get the Schroedinger equation $\psi''-U(z)\psi=0$ with
\[
U(z)=\frac{1}{2}A''+\frac{1}{4}A'^{2}-\gamma_{2}Re^{3A}.
\]

Now the resolution of the problem is translated in finding a square integrable solution to the above potential
\begin{equation}\label{condition}
\int dz\frac{1+\gamma_1 f}{1+\gamma_1 h}\psi^{2}=1,
\end{equation}
such that
\begin{equation}\label{gauge}
\int dz\frac{1+\gamma_1 f}{1+\gamma_1 h}\psi^{3}=\int dz\frac{1+\gamma_1 f}{1+\gamma_1 h}\psi^{4},
\end{equation}
leaving us with a consistent four dimensional theory.

Firstly, we can see that the only effect of the non-minimal coupling with the field strength is the changing of
the measure in the above integrals. However since for large $z$ we have that $h$ and $f$ are constant, the solution without the mass coupling $\psi=e^{A/2}$~\cite{Alencar:2014moa}
does not give a localized solution. This is consistent with the previous results of Refs.~\cite{Dvali:2000rx,Germani:2011cv}. However, as stressed before what will guarantee the full localized solution is the non-minimal coupling with the mass term. Since the equation of motion is not affected by the coupling with the field strength, the solution is obtained by fixing $\gamma_2=1/16$ with solution $\psi=ae^{A}$
as found in Ref.~\cite{Alencar:2014moa}, where $a$ is a normalization constant. This therefore can be plugged directly in Eq. (\ref{condition}). With this and after some manipulations we find a normalization constant given by

$$
c=(\frac{1+\gamma_{1}h/3}{1+\gamma_{1}h}\int e^{2A}dz)^{-1/2}
$$

which is convergent for any warp factor recovering RS for large $z$. Therefore our solution is localized and warp independent. The next problem to be solved is about the non-abelian gauge invariance in four dimensions. Since we yet have the parameter $\gamma_1$ and Eq. (\ref{gauge}) is a scalar equation this is in principle possible. In fact when we plug our solution in Eq. (\ref{gauge}), after some manipulations we get
\begin{equation}
\gamma_{1}=-\frac{\int(\psi^{3}-\psi^{4})dz}{\int f(\psi^{3}-\psi^{4})dz}.
\end{equation}

The above integral is also convergent. The numerator is trivially convergent since for large $z$ it converges faster than (\ref{condition}). The denominator is weighted by $f$ which is a constant for large $z$ and therefore it is convergent for any warp factor. Therefore the problem is completely solved and warp independent. The above results are also valid for the RS1 model. This is due to the fact that our analytical solution is integrable for any range of the integration in the extra dimension sector. The case of RS1 is particularly interesting. In Ref. \cite{Batell:2005wa}
 the model of Ref. \cite{Ghoroku:2001zu}, with boundary and mass term, has been used to analyze phenomenological consequences of localizing gauge fields. Since with this model the relation between boundary and mass term is not fixed, the authors can choose to localize the zero mode of the gauge field in either brane. In our model the solution to the zero mode is completely fixed to $e^{A}$ and therefore we can only have gauge fields localized in the UV brane. The holographic interpretation of such result has also been clarified in Ref \cite{Batell:2005wa}. The authors showed that when the mode is localized on the UV brane, the photon eingenstate in the dual theory is primarily composed of the source field.

As concluding remarks we should point out that our model solves the long
standing problem of consistently obtaining non-abelian YM fields in RS scenarios from a bulk action. The strong points
are that it solves the problem for arbitrary warp factors
beyond does not adding any other fields or degrees of freedom. Although
non-minimal couplings with gravity has not been much exploited in
RS models the resolution of this problem poses this kind of couplings
in a central position. Many questions arise such as the origin of this coupling or
if there are other couplings which can solve the problem. If this
is true we should also answer what kind of restriction must be imposed
to get a desirable theory in four dimensions. In fact in separate works we have analyzed resonances of the $U(1)$ model\cite{Jardim:2015vua}. The results point that it does not possess resonances of massive modes, what seems at least curious. Soon later we generalized the case of $U(1)$ gauge field to include interactions with other geometrical objects\cite{Alencar:2015oka}. The result is that there are some kinds of tensors which do not provide a localized zero mode. In these cases we also do not found resonances for the cases considered. From another viewpoint, the fact that our five dimensional action is not gauge invariant points to the possibility of considering other interactions which break this symmetry.  Another question is if these kind of couplings work for other fields. In this direction some of us has shown
that the non-minimal coupling with the mass term can be used to localize
ELKO spinors and $p-$form fields in a warp independent way~\cite{Alencar:2014moa,Alencar:2014fga,Jardim:2014cya,Jardim:2014xla}. Despite the fact that the problem seems to be punctual, the way used to solve it points to many interesting directions. It is clear for us that a wide range of possibilities has been opened. It is also important to understand the reason why these kind of couplings generate solutions which solve the problems independently of a specific form for the warp factor.
In particular, spin 1/2 fields are specifically interesting, but this
seems to be an yet more difficult problem since basically all the
models and extensions of RS give only one chirality of the field localized on the brane.
However, since our model provides a localized gauge field we must
also have a localized current, what seems to imply that the fermion field is also localized. However this is very speculative, since in the present model
we have not specified if the SM fields are localized near the IR, or if SM fields are allowed to propagate in the bulk (specially fermions). Possibly localized gauge field might be problematic in this context due to FCNC constraints. We should also point that the presence of two fine tuned parameters maybe a shortcoming that can be solved if more symmetries are added to the model, such as supersymmetry\cite{Gherghetta:2000qt}. Finally, in another direction, the coupling to the Ricci scalar with gauge fields will change the phenomenology of the RS model. The present model changes the coupling between the radion graviscalar and the gauge fields and this can be phenomenologically relevant at the LHC and also to cosmology of RS models\cite{Chacko:2013dra,Csaki:1999mp}. All the above points are under consideration by the present authors.

\section*{Acknowledgment}
Geova Alencar would like to thanks Moreira, A. A. for fruitful discussions. We also thank the referee for very fruitful criticism of the manuscript.  We acknowledge the financial support provided by Funda\c c\~ao Cearense de Apoio ao Desenvolvimento Cient\'\i fico e Tecnol\'ogico (FUNCAP), the Conselho Nacional de Desenvolvimento Cient\'\i fico e Tecnol\'ogico (CNPq) and FUNCAP/CNPq/PRONEX.

\providecommand{\href}[2]{#2}\begingroup\raggedright\endgroup

\end{document}